\documentclass[aps,prl,superscriptaddress,twocolumn,amssymb]{revtex4-1}
\usepackage{amsmath}
\usepackage{graphicx}
\usepackage{braket}
\usepackage{hyperref}

\begin{document}
\title{In situ characterization of qubit control lines: a qubit as a vector network analyzer}

\author{Markus Jerger}
\affiliation{ARC Centre of Excellence for Engineered Quantum Systems, The University of Queensland, St Lucia QLD 4072, Australia}

\author{Anatoly Kulikov}
\affiliation{ARC Centre of Excellence for Engineered Quantum Systems, The University of Queensland, St Lucia QLD 4072, Australia}
\affiliation{School of Mathematics and Physics, University of Queensland, Brisbane, Queensland 4072, Australia}

\author{Z\'enon Vasselin}
\affiliation{ARC Centre of Excellence for Engineered Quantum Systems, The University of Queensland, St Lucia QLD 4072, Australia}

\author{Arkady Fedorov}
\email{a.fedorov@uq.edu.au}
\affiliation{ARC Centre of Excellence for Engineered Quantum Systems, The University of Queensland, St Lucia QLD 4072, Australia}
\affiliation{School of Mathematics and Physics, University of Queensland, Brisbane, Queensland 4072, Australia}

\begin{abstract}
We propose and experimentally realize a technique to measure the transfer function of a control line in the frequency domain using a qubit as a vector network analyzer. Our method requires coupling the line under test to the longitudinal component of the Hamiltonian of the qubit and the ability to induce Rabi oscillations through simultaneous driving of the transversal component. The method can be used to increase the fidelity of entangling gates in a quantum processor. We have demonstrated that by characterizing the 'flux' control line of a superconducting Transmon qubit in the range from 1 to 450 MHz and using this characterization to improve the fidelity of an entangling CPHASE gate between two Transmon qubits.

\end{abstract}


\date{\today}

\maketitle
Signal distortions are inevitable in experiments involving radio frequency controls, where they can impact the quality of measurements and generate unwanted artifacts.
In quantum control experiments and quantum information processing these distortions are the source of errors and may limit the fidelity of operations.
Quantum gates that use non-adiabatic (fast) frequency tuning of the qubits involved are particularly sensitive to distortion and require precise calibration~\cite{Hofheinz2009,Bylander2009,Johnson2011PhD, Bozyigit2011, Baur2012b,Langford2016}.
Distortion can be canceled, in principle, if the complex transfer function of the control line is known, by applying its inverse to the signal before it is transmitted.
The most common approach to obtain the transfer function is to measure it (at room temperature) in the frequency domain using a vector network analyzer or in the time domain using an oscilloscope (see, for example, Ref.~\cite{Baur2012b}).


This method has two important deficiencies: the transfer function of the line changes when the setup is cooled to cryogenic temperatures, and the part of the signal line from the microwave connector closest to the chip to the qubit is not included in the characterization. Various methods for {\it in situ} line calibration have been proposed. Some calibration methods are limited in time resolution by the length of the microwave $\pi$-pulse~\cite{Hofheinz2009,Johnson2011PhD,Baur2012b}, others are applicable only to specific systems~\cite{Bylander2009} or pulses~\cite{Gustavsson2013}, and most procedures only provide indirect information about the transfer function. 

In this Letter we propose and experimentally realize a method of {\it in situ} direct reconstruction of the response of a control line of a qubit in a large frequency range using the qubit itself. We benchmark the method by measuring the transfer function of an element introduced at room temperature and comparing the measured response with the response obtained by using a commercial vector network analyzer (VNA). We then apply the method to improve the fidelity of a non-adiabatic controlled phase (CPHASE) gate \cite{Strauch2003, DiCarlo2009} between two Transmon qubits, one of the most commonly used entangling gates in superconducting systems. 



To understand the principles underlying our method consider the Hamiltonian of a qubit with time dependent longitudinal (frequency control) and transversal (excitation) drives
\begin{align}\label{H}
	H &= \frac{\hbar}{2} \omega_0 \sigma_z + \hbar A_x \cos \left( \omega_x t + \phi_x \right) \sigma_x \nonumber \\
		&+ \hbar A_z \cos \left( \omega_z t + \phi_z \right) \sigma_z. 
\end{align}
After transformation into the rotating frame  with $U_1 = e^{i (\omega_x t + \phi_x) \sigma_z / 2}$, the Hamiltonian reads
\begin{equation}  \label{1stH}
	H' = \frac{\hbar}{2}A_x \sigma'_x + \frac{\hbar}{2} \delta\omega \sigma'_z 
	   + \hbar A_z \cos \left( \omega_z t + \phi_z \right) \sigma'_z,
\end{equation}
where $\delta\omega = \omega_0 - \omega_x$ is the detuning of the excitation driving and we have used the rotating wave approximation under the assumption $A_x \ll \omega_0$. Note that the transversal phase $\phi_x$ does not explicitly appear in the Hamiltonian (\ref{1stH}), however it implicitly provides a reference for the rotating frame through the transformation operator $U_1$. The time-independent part of $H'$ can be diagonalized with $U_2 = e^{i \phi_t \sigma'_y/2}$, $\phi_t = \arctan \left(\delta \omega / A_x \right)$, as
\begin{align}\label{1stH diag}
    \tilde{H}' &= \frac{\hbar}{2} \Omega_R \tilde{\sigma}'_x + \hbar\omega_R \cos \left( \omega_z t + \phi_z \right) \tilde{\sigma}'_z \nonumber 
    \\
		&+ \hbar\omega_R\frac{\delta\omega}{A_x} \cos \left( \omega_z t + \phi_z \right) \tilde{\sigma}'_x,
\end{align}
where $\Omega_R = \sqrt{A_x^2 + \delta \omega^2}$ is the Rabi frequency and $\omega_R = A_z A_x/\Omega_R$. 
In the special case of $\delta\omega = 0$, $U_2$ reduces to the identity, $\Omega_R = A_x$ and $\omega_R = A_z$.

\begin{figure*}[t]
	\begin{center}
		\includegraphics[width=\textwidth]{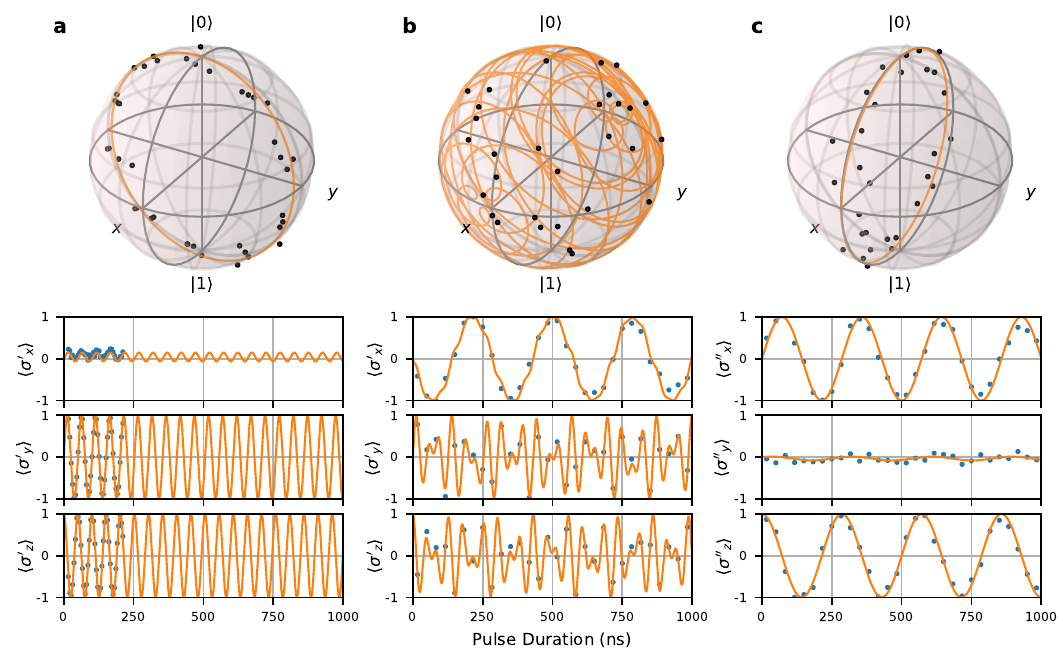}
		\caption{
			A typical data set for $\Omega_R = 19.8$~MHz and the corresponding trajectories of the Bloch vector in the first rotating frame with  only $x$ drive applied, with both $x$ and $z$ drives and with both $x$ and $z$ drives in the second rotating frame. 	
			a) 
				State tomography of the qubit with only the $x$ drive applied. 
				The fit yields a rotation vector of $\vec{\Theta}$=(19.7,-1.9,0.8)~MHz and $\phi_x = -0.1$.
			b) 
				State tomography of the qubit with both the $x$ and $z$ drives applied. 
				The frequency of $z$ drive is set to $\omega_z = \Omega_R$. 
				The sampling rate of the experiment is chosen such that the slow oscillations at $\omega_R$ can be resolved unambiguously but not necessarily the fast oscillations at $\Omega_R$, which are canceled by $U_3$.
				The orange curve shows the theoretical dynamics derived from the fit in (c). 
			c) 
				The dynamics of the qubit in the second rotating frame, transformed from (b). 
				The fit yields $\vec{\theta}$=(0.2,-3.5,0.1)~MHz and we can obtain $A_z = 3.51$~MHz and $\phi_z = -1.54$.
		} 
		\label{fig:trace}
	\end{center}
\end{figure*}

Comparing (\ref{H}) and (\ref{1stH diag}), we observe that in the rotating frame the $z$ term plays the role of a transversal drive for the dressed-state qubit with splitting $\Omega_R$ and will induce Rabi oscillations of the dressed-state qubit with frequency $\omega_R$. To make this observation explicit we transform the Hamiltonian ($\ref*{1stH diag}$) into the second rotating frame for $\omega_z = \Omega_R$ with $U_3 = e^{i (\omega_z t) \tilde{\sigma}'_x / 2}$ to obtain
\begin{equation}\label{H2d}
	H'' = \frac{\hbar}{2}\omega_R  \left( \sin \phi_z \, \sigma''_y + \cos \phi_z \, \sigma''_z \right),
\end{equation}
where we have used another rotating wave approximation with $\omega_R \ll \Omega_R$. The Hamiltonian $H''$ shows that the amplitude $A_z$ and phase $\phi_z$ of the $z$ control are encoded in the frequency and axis of the Rabi oscillations of the dressed-state qubit and can be measured in the experiment.

By setting $\delta\omega = 0$ and varying $A_x$ one can perform the experiment for different $\omega_z$ and identify the transfer function of the $z$ line in the range of frequencies $\Gamma_1, \Gamma_2 \ll \omega_z \ll \omega_0$ which are most relevant for the frequency control of the qubit with  $\Gamma_{1,2}$ being the relaxation and dephasing rates of the qubit, respectively. 

For weakly anharmonic qubits, such as the Transmon, the two-level approximation breaks down when $A_x$ starts being a considerable fraction of the anharmonicity, about 350~MHz for the qubits used here. To mitigate this problem, we can exploit off-resonant driving ($\delta \omega < 0$), which increases the $\Omega_R$ at the same $A_x$ and also allows higher $A_x$ because the frequency of the drive is further from the frequency of the $\ket{1}$-$\ket{2}$ transition. This allows us to extend the analysis to higher frequencies at the expense of signal amplitude.

Both $A_z$ and phase $\phi_z$ for a given frequency $\omega_z$ can be determined directly from the observables in the laboratory or the first rotating frames, such as the excited state population of the qubit. However, it is more convenient to perform tomography and reconstruct the oscillation of the qubit state in the second rotating frame by post-processing. Removing the fast population oscillation with frequency $\Omega_R$ and leaving only the signal varying with the frequency $\omega_R$ allows for a substantial reduction of the required sampling rate and more robust fitting. The experimental procedure is summarized by the following steps.
\begin{itemize}
	\item Apply $x$ drive with $z$ drive off ($A_z =0$) and fit data to extract $\Omega_{R}$ (Fig.~\ref{fig:trace}a).
	\item Set $\omega_z = \Omega_{\rm R}$, apply both $x$ and $z$ drives and use tomography pulses to reconstruct $\langle \sigma'_x(t) \rangle$, $\langle \sigma'_y(t) \rangle$ and $\langle \sigma'_z(t) \rangle$ in the first rotating frame (Fig.~\ref{fig:trace}b).
	\item Post-process the data to reconstruct $\langle \sigma''_x(t) \rangle$, $\langle \sigma''_y(t) \rangle$ and $\langle \sigma''_z(t) \rangle$ in the second rotating frame. Fit the resulting data to Rabi oscillations described by (\ref{H2d}) to extract $A_z$ and $\phi_z$ for given  $\omega_z$ (Fig.~\ref{fig:trace}c).
	\item Repeat the sequence for different $\Omega_R$ to cover the necessary frequency range.
\end{itemize}


We describe the experimental steps in more detail and directions for efficient implementation of the procedure in Appendix 1. We have tested our method on a standard circuit quantum electrodynamics system: a Transmon qubit coupled to a readout resonator with local charge and flux lines. More specifically, we choose the QB~2 of a chip virtually identical to one used in Ref.~\cite{Steffen2013a}.

\begin{figure}[tb]
	\begin{center}
		\includegraphics[width=\columnwidth]{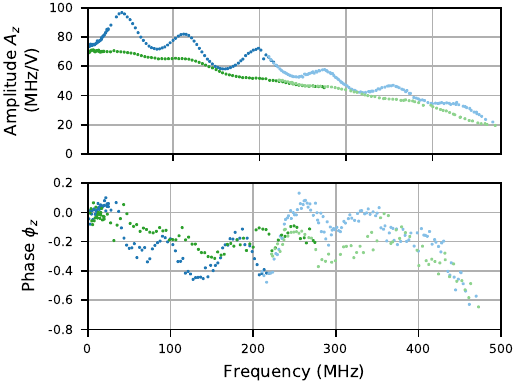}
		\caption{
			Amplitudes $A_z$ and phases $\phi_z$ of the transfer functions of the $z$ control line. 
			Points in blue and green show the response of the flux line for two different configurations (see text for details). Darker points were measured with resonant driving $\omega_x = \omega_0$ by varying $A_x$, lighter points were measured by varying $\omega_x - \omega_0$ with fixed $A_x$ (off-resonant case).
		}
		\label{fig:amplitude_vs_offset}
	\end{center}
\end{figure}

\begin{figure}[ht]
	\begin{center}
		\includegraphics[width=\columnwidth]{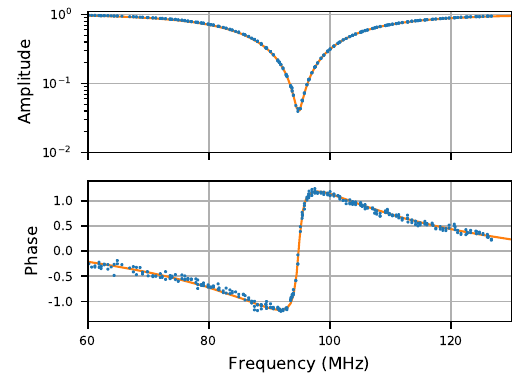}
		\caption{
			Amplitude and phase of transmission through a transmission line with a shorted stub resonator.
			Points in blue were measured with the qubit by comparing $A_z$ and $\phi_z$ with and without the stub in place, the orange line was measured directly with a commercial vector network analyzer.
		}
		\label{fig:qubit_vs_vna}
	\end{center}
\end{figure}

With our method we characterized the complex transfer function in the range of 1 to 450\,MHz 
(Fig.~\ref{fig:amplitude_vs_offset}). Each point was taken with 4,096 averages at a repetition rate of 40\,kHz. 
The measurements at lower frequencies were limited by decoherence of the qubit. However, we point out that the low frequency part of the transfer function can be also measured with other methods with lower time resolution~\cite{Hofheinz2009,Johnson2011PhD}. 

At high frequencies our accuracy is limited by population of the higher levels for the resonant driving case and by the loss of signal contrast at large detunings in the off-resonant driving case. In addition, the amplitude reconstruction is more robust as it corresponds to the frequency of the oscillations while the phase is reconstructed from a ratio of their amplitudes. 

We have benchmarked our method by introducing an additional element in the flux line at the room temperature. We used a shorted stub resonator made from a BNC T-adapter and several meters of a BNC cable shorted at the end. We repeated the characterization of the line with the element and used the original data for the line to de-embed the transfer function of the element itself. The result is shown in Fig.~\ref{fig:qubit_vs_vna} in comparison with the transfer function of the element measured separately with the commercial vector network analyzer (VNA). The agreement between our method and VNA is excellent, showing the dynamic range of our method for measuring amplitude of $\simeq 30$~dB. The dynamic range of a single measurement is bounded by the condition $\omega_R \ll \Omega_R$ required for the rotating wave approximation in (\ref{H2d}) and the decoherence time of the qubit. It can be further improved by dynamically changing the amplitude of the $z$ drive, taking advantage of the dynamic range of the AWG.


We have employed the outlined method of flux line calibration to improve the quality of a CPHASE entangling gate  between two superconducting transmon qubits. The gate between two transmons can be realized~\cite{Strauch2003, DiCarlo2009} by bringing the second excited state $\ket{02}$ of one qubit in resonance with the $\ket{11}$ state where both qubits are excited, inducing energy exchange between the states. On each swap of an excitation the state acquires a phase of $\pi/2$, so that a system starting in $\ket{11}$ evolves into $i \ket{02}$ and subsequently into $-\ket{11}$. Since one of the qubits must be tuned in frequency, there is an additional dynamical phase acquired by the qubit which must be accounted for. Apart from that, the other computational basis states, $\ket{00}$, $\ket{10}$ and $\ket{01}$ are unaffected by the gate.

\begin{figure}[tb]
	\begin{center}
		\includegraphics[width=\columnwidth]{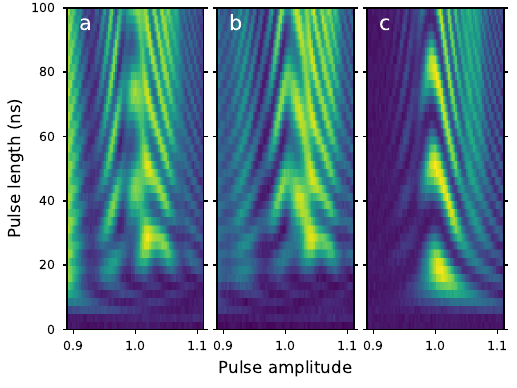}
		\caption{
		    Vaccum Rabi oscillation between the $\ket{11}$ and $\ket{20}$ states of two Transmon qubits assuming (a) a perfect impulse response, (b) the impulse response measured at room temperature and (c) the impulse response measured using the qubit.
		}
		\label{fig:chevrons}
	\end{center}
\end{figure}

To implement the gate we used QB1 and QB2 of the chip both parked at their symmetry points at frequencies $5.5$~GHz and $6.5$~GHz, respectively. We tune QB2 near $\sim5.8$~GHz to allow the states $\ket{11}$ and $\ket{02}$ to exchange excitation via virtual photons in a common transmission line resonator.

An ideal CPHASE gate requires a perfect step pulse, which cannot be implemented by any physical device. In order to comply with bandwidth limitation of our AWG and to make the distortion correction easier, we applied a Gaussian low-pass filter with a cut-off frequency of $300$~MHz to the ideal square pulse shape. This pulse shape was programmed to the AWG and was used to implement the CHPASE gate.

To calibrate the amplitude and duration of the flux pulse we prepared qubits in $\ket{11}$ followed by the flux pulse on QB2 where we sweep both the length and the amplitude of the pulse. Measurement of the resonator response shows characteristic oscillations (often called 'chevron patterns') manifesting excitation exchange between $\vert11\rangle$ and $\vert02\rangle$ states of the two transmons. The pattern is distorted due to extra reflections on the flux line which appear only at low temperatures and cannot be calibrated at room temperatures (the reflections of the line can be seen as broad resonances in the line frequency response, see Fig.~\ref{fig:chevrons}a). 

Following the most common approach to compensate for these distortions, we have measured the room temperature (RT) impulse response of the flux line before cooling down the refrigerator. We then used the standard machinery~\cite{Baur2012b} to compute the necessary waveform which then can be loaded to AWG to yield the desired pulse shape. Using this computed pulse shape, we repeated the calibration routine (Fig.~\ref{fig:chevrons}b) but the oscillations remain heavily distorted. In particular, one may note that the oscillations start only after the first fifteen nanoseconds, which we attribute to a large overshoot in the actual flux pulse. 

With our {\it in situ} calibration of the flux line (the response in blue on Fig.~\ref{fig:amplitude_vs_offset}) for distortion compensation, we have measured the improved oscillation pattern shown in Fig.~\ref{fig:chevrons}c. The oscillations visibility is asymmetric which we attribute mostly to the bandwidth limitations of the pulse shape. Apart from that the compensation fixes most of the imperfections and more specifically the oscillations begin without delay at short lengths of the flux pulse.

To quantify our method, we have calibrated the CPHASE gate with no correction, with RT correction and our {\it in situ} correction. Applying our method the process fidelity of the gate increases from 0.835$\pm 0.015$ to 0.875$\pm 0.01$. Numerical simulation of the protocol with the experimental values of $T_1=2.5~\mathrm{\mu s}$ and $T_2=1.4~\mathrm{\mu s}$ and perfect pulse shape showed the process fidelity of 0.885. Contrary to our initial expectations, applying the RT calibration has not improved the fidelity of the gate.

During a later cooldown we introduced extra attenuation to the resonator input and flux lines which increased the coherence of the qubits to $T_1=4~\mathrm{\mu s}$, $T_2=4.5~\mathrm{\mu s}$. {\it In situ} measurement of the frequency response of the line (Fig.~\ref{fig:amplitude_vs_offset}, green curves, rescaled for the extra attenuation) did not show the characteristic resonances seen previously (Fig.~\ref{fig:amplitude_vs_offset}, blue curves). In this configuration we were unable to demonstrate statistically significant improvement in the fidelity of the CPHASE gate either with RT-calibration or with {\it in situ} calibration. Process fidelity was measured to be 0.945, which was predominately limited by decoherence of the transmons. 

Our method is the first direct {\it in situ} measurement of the line transfer function from room temperature electronics to a qubit on a chip. The method is most relevant for superconducting qubits whose frequencies are routinely tuned but is applicable for all qubits with $z$ and $x$ control. For superconducting qubits one can use our procedure to improve the fidelity of the two-qubit quantum gates~\cite{Strauch2003, DiCarlo2009, Bialczak2010, Yamamoto2010} as well as photon-qubit operations requiring non-adiabatic control~\cite{Hofheinz2009,Bozyigit2011,Langford2016}. In addition to quantum control applications the qubit can be also used as a microscopic probe of the electromagnetic fields in frequency domain. 

\begin{acknowledgments}
	 We thank Steffen Schl\"or for his help with the experiment. MJ, AK and AF were supported by the Australian Research Council Centre of Excellence CE110001013. AF was supported by the ARC Future Fellowship FT140100338.
\end{acknowledgments}

\bibliographystyle{apsrev4-1}
%

\newpage

\  \

\newpage

\section{Appendix 1}

\ 

In this section we go through the experiment step-by-step, giving some directions for efficient implementation of the procedure. We also discuss some of the complications we have encountered performing the measurements and ways to mitigate them.

{\it Step 0:} Make sure (\ref{H}) is a good approximation of the physical system. In case of the Transmon and other superconducting qubits, that means choosing a magnetic flux bias and offset (longitudinal driving) amplitude that result in a predominantly linear dependence of the qubit transition frequency on the offset.

{\it Step 1:} Apply an $x$ drive to the qubit with the $z$ drive off. The $x$ drive can use square pulses or any pulse shape with an envelope that has a constant amplitude section. Sweep the duration of the constant amplitude section to observe Rabi oscillations. Perform tomography to reconstruct the evolution of a state vector $\vec{v}(t) = \left( \langle \sigma'_x(t) \rangle, \langle \sigma'_y(t) \rangle, \langle \sigma'_z(t) \rangle\right)$ in the first rotating frame (see Fig.~\ref{fig:trace}a). 

Fit $\exp(\vec{\Theta}_R t \cdot \vec{L}) \vec{v}(0)$ to this data with fitting parameters $\vec{\Theta}_R = \left( \Theta_x, \Theta_y,\Theta_z\right)$. Here, $t$ is the duration of the constant amplitude section of the pulses and $\vec{L} = \left(L_x, L_y, L_z\right)$, $L_x = \left( \begin{smallmatrix}0 & 0 & 0\\ 0 & 0 & -1\\0 & 1 & 0\end{smallmatrix}\right)$, $L_y = \left( \begin{smallmatrix}0 & 0 & 1\\ 0 & 0 & 0\\-1 & 0 & 0\end{smallmatrix}\right)$, $L_z = \left( \begin{smallmatrix}0 & -1 & 0\\ 1 & 0 & 0\\0 & 0 & 0\end{smallmatrix}\right)$ are the SO(3) generators of rotations about the $x$, $y$ and $z$ axis, respectively, and  $\vec{v}(0) = \left(0, 0, -1\right)$ is the initial state of the system. The direction of $\vec{\Theta}$ determines the axis of rotation of the state vector on the Bloch sphere and the absolute value gives the Rabi frequency $|\vec{\Theta}|=\Omega_R$. If non-square pulses are used, $\vec{v}(0)$ can be made a fitting parameter to absorb the attack of the pulses.

Ideally we expect $\vec{\Theta} = \left( A_x, 0,\delta \omega\right)$. However in the experiment we often observed small spontaneous detuning of the qubit $\Theta_z\neq0$ at larger amplitudes $A_x$.  As an example, the data for $\Omega_R = 19.8$~MHz was taken assuming $\delta\omega = 0$ but  shows (see Fig.~\ref{fig:trace}a) small oscillations for $\langle \sigma'_x(t)\rangle$ component. This spontaneous detuning can be absorbed in $\delta\omega$ and does not affect the results of the line calibration.

In addition, we also observed a spurious component $\Theta_y\neq0$ due to phase differences between the tomography pulses and the qubit driving pulse. In particular, for $\delta\omega \neq 0$ the frequency of the driving pulse is different from the frequency of the resonant tomography pulses which may lead to some phase difference depending on technical details of the pulse generation. This effect can be canceled in the post-processing of the data in Step 3.

{\it Step 2:} Apply $x$-drive as in Step 1, add simultaneous $z$-drive at frequency $\omega_z = |\vec{\Theta}|$. Use a square pulse envelope for the $z$ driving pulses and sweep the duration of the $x$ and $z$ pulses.  Use tomography to reconstruct the evolution of a state vector $\vec{v}_0(t) = \left( \langle \sigma'_x(t) \rangle, \langle \sigma'_y(t) \rangle,\langle \sigma'_z(t)\rangle\right)$. Due to $x$-drive the $y$ and $z$ components of the state vector oscillate with the Rabi frequency $\Omega_R$ while the $z$-drive modulates these oscillations at $\omega_R\ll\Omega_R$ (see Fig.~\ref{fig:trace}b). The duration of the pulses should be long enough to observe these slower oscillations.

{\it Step 3:} Post-process the data to reconstruct the state vector rotation in the second rotating frame. That step has the benefit of removing the fast Rabi oscillations, thus allowing to take fewer measurements, and allows to fit simpler expressions with less parameters to the data. 

To post-process the data we first correct for the phase shift between the $x$ driving pulse and tomography pulses by applying the transformation $\vec{v}_1(t) = \exp(\phi_x L_z)\vec{v}_0(t)$, where $\phi_x = \arctan \left(\Theta_y/\Theta_x\right)$. To account for the transformation from (\ref{1stH}) to (\ref{1stH diag}) we rotate the basis $\vec{v}_2(t) = \exp(\phi_d L_y)\vec{v}_1(t)$, where $\phi_d =-\arctan \left[\Theta_z/\left(\Theta^2_x+\Theta_y^2\right)^{1/2}\right]$. Finally we remove the fast oscillations by going to the second rotating frame  $\vec{v}_3(t) = \exp(|\vec{\Theta}| L_x t)\vec{v}_2(t)$.

The resulting Rabi oscillations of the state vector in the second rotating frame described by $\vec{v}_3(t)= \left( \langle \sigma"_x(t) \rangle, \langle \sigma"_y(t) \rangle,\langle \sigma"'_z(t)\rangle\right)$ (see Fig.~\ref{fig:trace}c) are fit to $\exp(\vec{\theta}_R \cdot \vec{L}) \vec{v}(0)$, where both $\vec{\theta}=(\theta_x,\theta_y,\theta_z) $ and $\vec{v}(0)$ are fit parameters. The amplitude of the $z$-drive can be found as $A_z = |\vec{\theta}| |\vec{\Theta}|/\left(\Theta^2_x+\Theta_y^2\right)^{1/2}$ and the phase of the drive is given by $\phi_z = \arctan(\theta_y/\theta_z)$. Step 3 completes the characterization of the $z$-line at the frequency $\omega_z = \Omega_R$ and the procedure can be repeated for another frequency $\Omega_R$ controlled by the appropriate choice of $A_x$ and $\omega_x$.

\section{Appendix 2}

\begin{figure}[t]
	\begin{center}
		\includegraphics[width=\columnwidth]{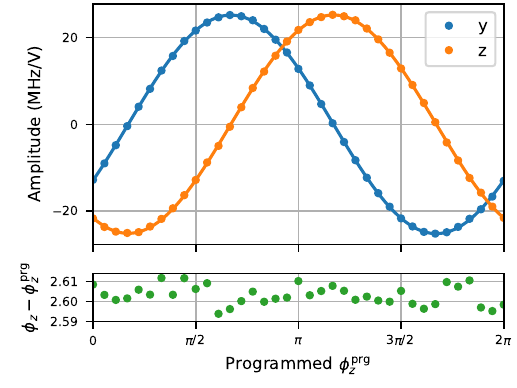}
		\caption{
			\emph{Quality of the phase reconstruction} -- 
			(top) Measured $y$ and $z$ components of the Rotation vector of the dressed qubit as the phase of the longitudinal driving pulse programmed at the AWG is varied.
			(bottom) Difference between the phase $\phi_z$ extracted by the method and the phase programmed at the AWG. 
		}
		\label{fig:phase}
	\end{center}
\end{figure}



The sensitivity of our method to the phase of $z$-drive is demonstrated in Fig.~\ref{fig:phase}. As expected a change in phase of the $z$-driving is reliably detected by our procedure. This should be contrasted to the conventional Rabi oscillations as only the amplitude of the drive can be accessed from the frequency of the oscillations while the phase is not accessible.


\end{document}